# ECR Plasma assisted deposition of zinc nanowires.

Vishwas S.Purohit[1], Manish Shinde[2], Shirshendu Dey[1], Renu Pasricha[3], C.V.Dharmadhikari[1], Tanay Seth[2] and S.V.Bhoraskar[1,*].

1) Department of Physics, University of Pune, Pune-411007, India
2) Centre for Materials for Electronics Technology, Pashan Road, Pune-411008., India
3) National Chemical Laboratory, Dr. Homi Bhabha Road, Pune 411 008, India

Abstract

Deposits of one-dimensional nanostructures of zinc with diameters of 90-120 nm have been obtained by means of sputtering in an Electron Cyclotron Resonance (ECR) plasma reactor. The sputtering has been made effective by using a negatively biased cylindrical target in a "hollow cathode" geometry. The nanocrystalline films deposited on the glass substrate were studied with scanning tunneling microscopy/spectroscopy and was found to consist of nanowires having metallic nature. The crystalline nature of these metallic nanowires was studied with x-ray diffraction analysis.

Corresponding author:- *svb@physics.unipune.ernet.in (S.V.Bhoraskar)

**Introduction: -**

Motivated by the fascinating electrical, catalytic and optical properties of the one-dimensional metal nano-structures [1, 2], many methods have evolved to synthesize these materials. Zinc is one of the most interesting materials for the fundamental research such as self assembled nanodomains and thermoelectric and magneto-resistance properties in nanowires [3, 4, 5]. One dimensional structures of zinc have been synthesized in number of ways [6] which include evaporation of ZnO powder mixed with graphite in an $NH_3$ flowing carrying gas environment [7], the reaction of milled ZnO powder with ammonia [8, 9], by evaporation of Zn grains in low vacuum at $200^oC$ [10] and by using vapor transport process on a zinc foil[11]. Here we report the use of microwave assisted plasma for depositing nanocrystalline zinc wires. Although applications of low pressure plasma devices for depositing varieties of thin films by chemical vapor deposition is well documented. There are only a few reports on its applications for depositing nanoparticles and one dimensional structures. [12,13,14].Microwave assisted Electron Cyclotron Resonance (ECR) low pressure plasma is one of these which is very popularly used for depositing and processing of thin films and surfaces [15,16].

In the present paper, nanowires of zinc were deposited on glass substrates after sputtering it from a cylindrical cathode in an ECR plasma. The deposits were then characterized by the techniques of Transmission Electron Microscopy (TEM), Scanning Electron Microscopy (SEM) and Scanning Tunneling Microscopy (STM) for understanding the structure and morphology. From the SEM micrographs, it was inferred that the growth is initiated by the particles which subsequently lead into nanowires. The nanowires were seen to exhibit metallic behavior as was confirmed by Scanning Tunneling Spectroscopic (STS) studies. The paper reports a simple method involving confined sputtering for producing one dimensional crystalline zinc nanowires by making use of an Electron Cyclotron Resonance (ECR) plasma.

**Experimental details:**

The plasma was excited by introducing hydrogen as a carrier gas into the electron cyclotron resonance (ECR) reactor chamber. The schematic of the experimental setup is shown in Fig. 1(a). The ECR plasma was excited by 2.45 GHz microwave source in $TE_{11}$ mode inside a resonance cavity along with the required magnetic field of 875 Gauss generated by electromagnets. The ECR cavity consisted of a cylindrical stainless steel chamber, 15 cm in height and 12.5cm in diameter and was coupled to the reactor chamber having a height of 30 cm and diameter of 20 cm. The reactor chamber was facilitated with various ports like gas inlet, vacuum port sample holder port and feed throughs. The 500W microwave source was coupled through a quartz window to the resonance cavity. Base pressure was $10^{-5}$ mbar and operating pressure was $10^{-2}$ mbar.

A hollow cylinder of zinc metal was mounted horizontally inside the processing chamber at a distance of 20 cm from the ECR zone and served as the sputtering target. The cylinder was 12 mm in diameter and 30 mm in length. Its axis was kept transverse to the main reactor axis. Rectangular glass substrates, $25 \times 10 mm^2$, were placed inside the cylinder as substrate for deposition. A negative bias voltage of 300V was applied to the target cylinder with reference to the chamber which was kept at the ground potential. The deposition time was varied from 10 to 30 minutes in three steps.

The glow from the plasma for hydrogen inside the hollow cylinder was seen to be brighter than that compared to the surroundings. The plasma density was estimated from the Langmuir double probe measurements carried out by keeping the probes inside the negatively biased cathode. Plasma density was calculated from the I-V characteristics following the standard procedure used in Langmuir double probe technique [17]. The electron density estimated from the probe currents, is plotted as a function of the negative bias voltage on the cylindrical cathode. The density is seen to increase with increasing voltage and is around six times higher at 300 V compared to the one obtained at no bias.

Scanning Electron Microscope [JEOL-JSM-6360A] was used to reveal the physical structure of the samples. STM and TEM analysis were used for much closer investigation of the growth morphology. A home-built scanning tunneling microscope based on a fine mechanical-screw-lever arrangement assembly with a compact four-quadrant three-dimensional scanner was used for STM investigation [18]. One of the most widely used test samples for STM is highly oriented pyrolytic graphite (HOPG) because of its reasonable flatness in the sub-nanometer range and non-reacting nature. The sample for STM analysis was prepared by putting a very dilute drop of the nanowire solution on the HOPG substrate followed by subsequent drying for 5 min. STM studies were carried out immediately after the deposition. The STM tips were made from 0.025 cm diameter polycrystalline Pt-Rh wires by mechanical cutting, at an angle. Details of the system and the general procedure for imaging are discussed elsewhere [19]. The images were processed using SPIP software [20].Scanning Tunneling Spectroscopy was employed for I-V measurement of the nanowires. XRD and Photoluminescence studies were also carried out.

**Results and Discussions**

The film deposited on the glass substrates placed inside the hollow cylindrical cathode assembly, was seen to be weakly adhesive to the substrate. The crystal structure and the crystalline phases were therefore studied by scrapping off the deposits and subjecting to X-ray diffraction technique. The XRD pattern is shown in Fig. 2 and is compared with the ASTM data for crystalline zinc. The sample does not seem to contain any impurity and is confirmed to adopt to hexagonal phase similar to its bulk counterpart. The average crystallite size was determined from the Debye Scherrer formula (particles size $t = 0.9 \lambda / \beta \cos \theta$ where $\lambda$ is the wavelength of x-rays, $\beta$ is the FWHM and $\theta$ is the Braggs angle) and was found to be 23 nm.

Scanning electron micrographs were recorded for the films deposited by sputtering for 10, 20 and 30 minutes and are presented in Fig 3. It is seen that in the initial stages, the films consisted of majority of nanoparticles, approximately 600nm in diameter as seen in Fig 3(a). By increasing the time of deposition to 20 minutes, nanowires are found to appear alongwith the particles. Whereas after 30 minutes of deposition the substrate is seen to completely get covered with nanowires. The particles have approximately 600nm diameter and the nanowires are 90-120 nm in diameter with average length of 10-20 μm.

TEM micrographs were recorded for the nanocrystalline powder, scrapped out from the substrate, supported by carbon-coated grids. Figs. 4(a), exhibits the micrograph and Fig 4(b) shows the Selective Area Diffraction (SAD) pattern. The surface of each nanowire is seen to consist of small particulates. The diameters of the nanowires are on an average seen to vary between 100-120 nm with length between 5-6 μm. Large number of Y joints are also seen. The interplanar separations were calculated from the SAD pattern for a single nanowire (observed under TEM) and it is found that the nanowire consist of both zinc and zinc oxide. The indices for metallic zinc which were identified include (100),(101), (002), (102) and (103). The plane (400) corresponds to zinc oxide. Thin oxide layer must have been grown on the zinc wire after exposure to atmosphere, since no signature is found for the oxide in the XRD spectrum.

The deposits were further characterized using scanning tunneling microscope with an optimum value of tunnel current of 0.1 nA and a bias voltage of 84mV in constant current mode. The time between the imaging and the sample preparation was kept to a minimum of 5 min. It was found that once the wires are located, with some difficulty they could be imaged repeatedly. I-V characteristic was also recorded on a single nanowire by interrupting the feedback loop.

Fig. 5(a) shows a 1000 nm x 1000 nm image. The nanowires are seen to be more than 1 μm in length and 120 nm in diameter as seen from the corresponding line scan shown in Fig. 5(b). These results are in good agreement with the value of around 120 nm obtained from TEM studies. Moreover, it appears that the surface of each nanowire is covered with much smaller (~5nm) structures. Imaging at predetermined small area of 100 nm × 100 nm [Fig. 5(c)] it is revealed that the surface of the nanowires are composed of still smaller nanowires of size 5-7 nm and nanoparticles of size 5 nm. Dendritic growth observed on the nanowires seen in the TEM image of Fig (4(a)) should have the similar structure. I-V measurements on these nanowires were recorded and typical result is shown in Fig 6. It exhibits linear behavior suggesting the metallic nature of the nanowires. However, we had observed the oxide structure from the analysis of SAD pattern. Under such a duality, it might be possible that the nanowires are metallic in nature wherein the nanoparticles get covered with thin oxide layer. The deposits of nanocrystalline Zn also exhibited room temperature photoluminescence with single broad peak centered at 408 nm as shown in Fig. 7. The luminescence from nanoparticles often arises on account of high concentrations of defect sites at surface regions of nanomaterials, which invariably get covered with an adsorbed layer of oxide. Presence of ZnO confirmed in the previous analysis is also supported by the luminescence.

The observed behavior of the deposit in the present experiments leading to the formation of nanostructures is interesting. Since such kind of product is not usually produced in the ECR plasma, the presence of negatively biased cylindrical sputtering target seems to be providing some unique characteristics. Such kind of geometry, known as "Hollow Cathode" is well documented by Koch et al[21] and Warner et al [22].

According to their explanations; in such geometry the electrons present in the plasma, are accelerated near the central glow region and move in a pendulum like motion. The to and fro motion of electrons cause increase in the ionization, thereby, leading to an increase in the density of hydrogen ions. Subsequently, hydrogen ions get accelerated towards the cathode to sputter out the zinc atoms. The neutral zinc atoms after sputtering out will get ionized due to the multiple collisions with electrons, and will be attracted towards the negatively charged substrate placed in the middle of the cylinder. The dominant forces responsible for particle transport seem to be electrostatic, neutral drag as well as self diffusion. The nucleation occurs after the deposition of zinc adatoms on the glass substrate, followed by the formation of chemical bonds. When the heat released from the formation of chemical bonds is sufficiently high and is not dissipated due to the poor thermal conductivity of the glass substrate, it disables further growth in the close proximity. This is known as island growth and has attracted considerable attention for a long time. It is also observed that such a phenomenon is strongly dependent on the material and operating conditions. [23,24]

However, further growth of the particle in the lateral direction to the substrate surface is possible. Apart from this the coalescence of smaller particles, in a preferred crystalline direction, is also possible. Both the processes will lead to the formation of nanowires. The sequence of nucleation and growth leading to the nanowires in the present experiments is illustrated schematically in Fig. 9.

From the observations of SEM micrographs presented in Fig 3(a) and (b) it is clear that the nanoparticles have grown into nanowires when the deposition time is increased from 10 mins to 20 mins. It is therefore clear that in the beginning individual particles are grown at each nucleating site. The coalescence and coagulation takes place in such a way that the surface free energy is minimized. The overall process can be considered to be the result of heterogeneous nucleation.

**Conclusions**

In conclusion, the hydrogen ion assisted hollow cathode microwave ECR plasma is an efficient technique for the growth of nanowires. The paper throws light on the metallic behavior of the nanowires using the scanning tunneling spectroscopic techniques. The high-resolution STM images have also helped us in understanding the microscopic morphology of the nanowires. From these studies, we conclude that the growth is a 3-step process in which the precursors are the particles, which consequently grow into fibers and wires. However, there seems to be complex phenomenon on account of the different kinds of charged and neutral species present in the plasma. The particles are seen to be covered with a thin oxide layer, which is inferred from the PL spectrum obtained from these metallic nanowires.

**Acknowledgements:-**

The Department of Science and Technology is acknowledged for the financial support.

**List of Figures Captions:** "Microwave ECR based synthesis of Zinc Nanowiers"

**Figure 1(a)** Schematic diagram of the hollow cathode sputtering assembly mounted inside the ECR plasma reactor at a distance of 20 cm from the ECR zone. **(b):** The I-V characteristic for hydrogen plasma by Double probe method.

**Figure 2.** X-Ray Diffraction pattern for the as prepared sample. Inset shows the JCPDS data for zinc metal.

**Figure 3.** SEM micrographs of sputter deposited zinc for **(a)** 10 minutes **(b)** 20 Minutes and **(c)** 30 minutes. The morphology is seen to change from nano particles to nanowires with increasing time of deposition.

**Figure 4.** Transmission electron micrographs of the zinc samples corresponding to 30 minutes deposition time.

**(a)** Image showing a serpentine geometry. Smaller nanoparticles, with 8-10 nm diameter, seem to be sticking as offshoots on the nanowires.

**(b)** Magnified image of the nanowires showing the particulates agglomerating on their surfaces

**(c)** Magnified image of a single nanowire with a capping layer on the surface

**(d)** SEAD Pattern for a single nanowires.

**Figure 5.** STM images of zinc nanothreads.

(A) Image recorded at lower magnification with constant-current mode for a scan-area of 1000 nm x 1000 nm. The parameters were $I_t$= 0.06 nA and $V_{bias}$ = 80 mV; White lines indicates the direction along with the Z profile plot is recorded and is exhibited in part B.

(B) Z profile plot corresponding to (A),

(C) Image recorded at higher magnification for a scan area of 100 nm x 100 nm. White lines indicates the direction along with the Z profile plot is recorded and is exhibited in part D

(D) Z profile plot corresponding to (C)

**Figure 6.** I-V characteristics of a single zinc nanowire recorded by scanning tunneling spectroscopy.

**Figure 7.** Room temperature photo luminescence spectrum of the as prepared zinc nanowires.

**Figure 8.** Schematic diagram explaining the sputtering process inside the hollow cathode.

**Figure 9**: Nucleation and growth of zinc particulates into nanowires on thermally insulating substrate like glass. (a) Adhesion (b) Nucleation (c) Cluster growth (d) Coalescence (e) Growth of nanowires

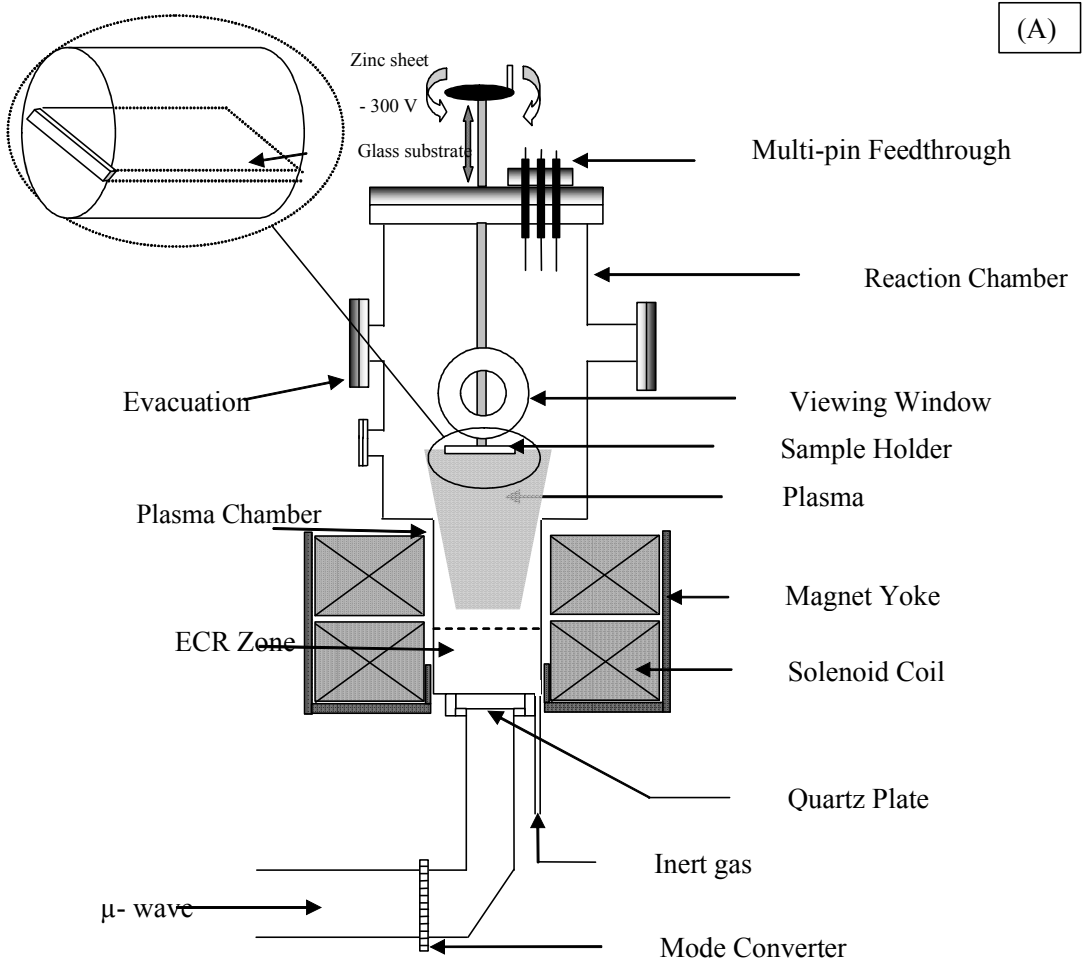
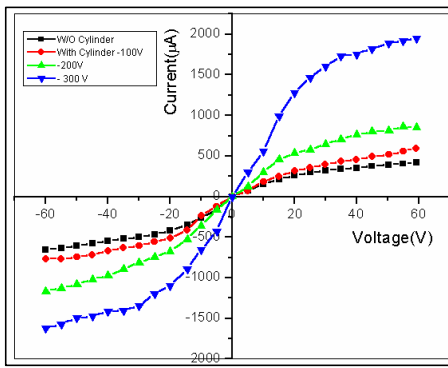
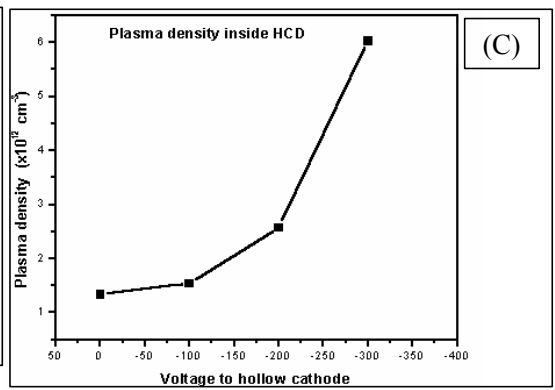

**Figure 1.**

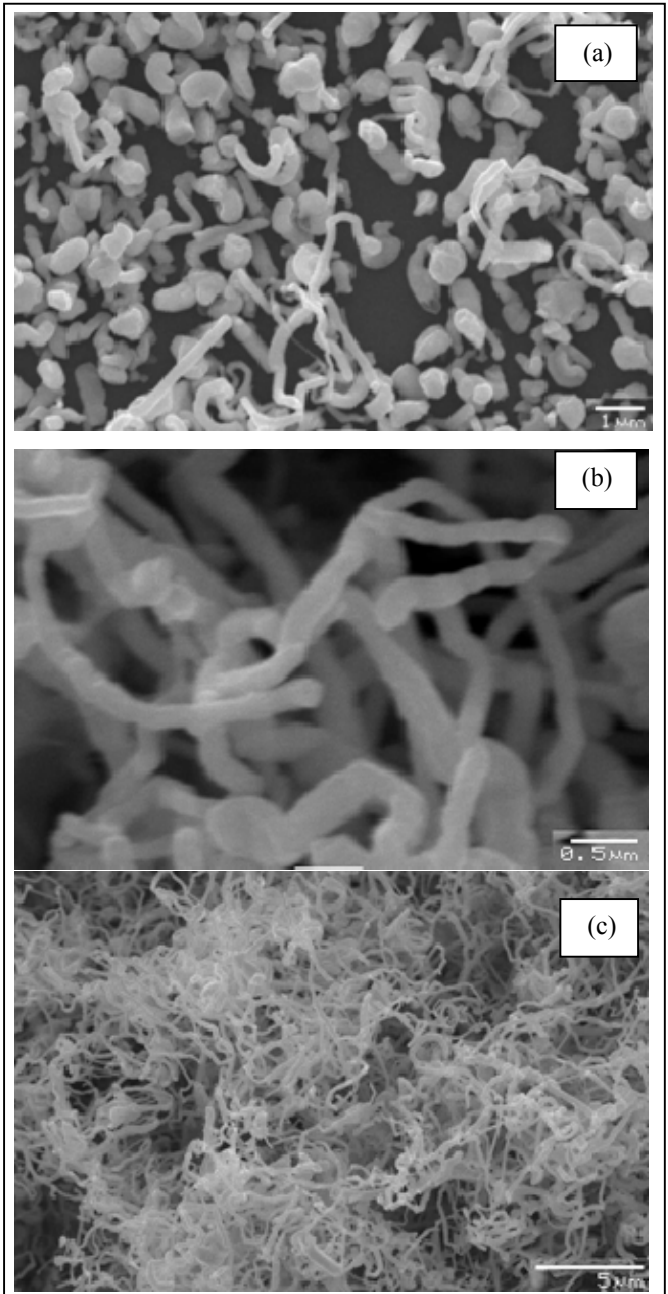

Figure 3

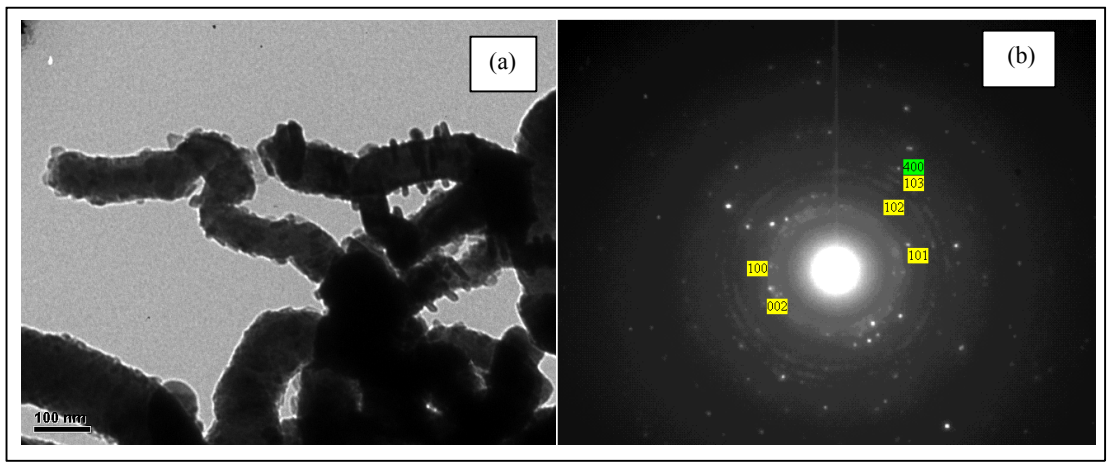

**Figure 4**

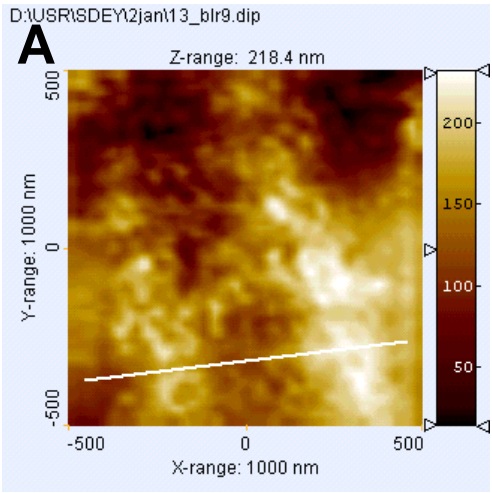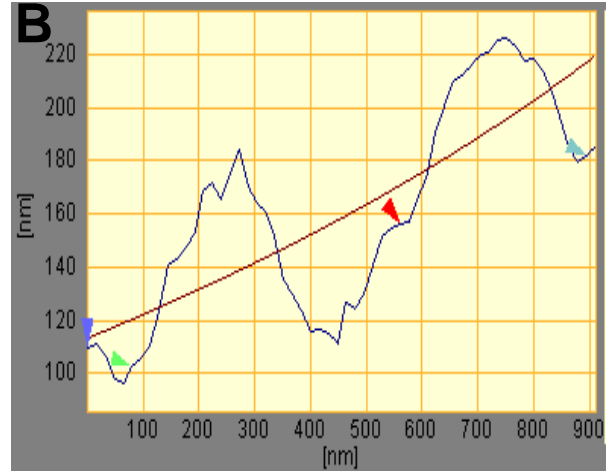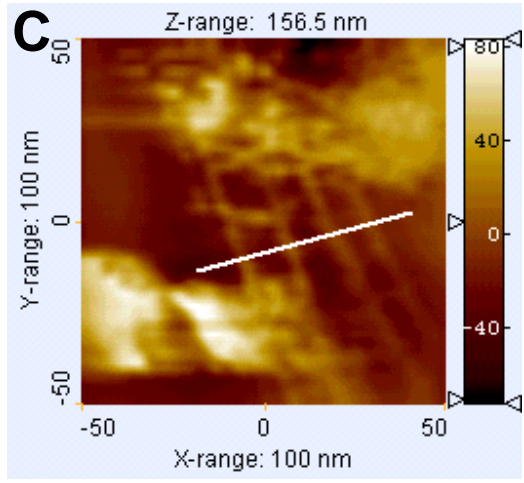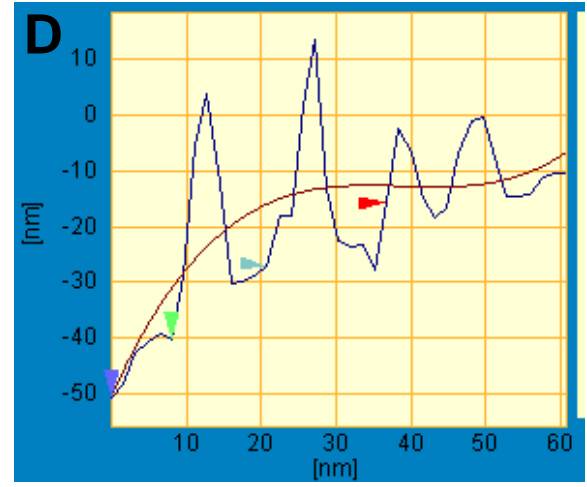

**Figure 5.**